\documentclass[prl,twocolumn,preprintnumbers,amsmath,amssymb,superscriptaddress]{revtex4}
\usepackage{epsfig}
\usepackage{amsmath}
\usepackage{graphicx}
\usepackage{esint}
\usepackage{siunitx}
\usepackage{color}

\usepackage{ulem}

\newcommand{\heriotwatt}{Institute of Photonics and Quantum Sciences, SUPA, Heriot-Watt University, Edinburgh EH14 4AS, UK}
\newcommand{\TsukubaKenji}{Research Center for Electronic and Optical Materials, National Institute for Materials Science, 1-1 Namiki, Tsukuba 305-0044, Japan}
\newcommand{\TsukubaTakashi}{Research Center for Materials Nanoarchitectonics, National Institute for Materials Science,  1-1 Namiki, Tsukuba 305-0044, Japan}

\newcommand{\icmuv}{Institute of Materials Science (ICMUV), University of Valencia, Valencia, Spain}

\begin{document}

\title{Photoexcitation of moir{\'e}-trapped interlayer excitons via chiral phonons}

\author{Antoine Borel}
\affiliation{\heriotwatt}

\author{Tatyana V. Ivanova}
\affiliation{\heriotwatt}

\author{Jorge Cervantes-Villanueva}
\affiliation{\icmuv}

\author{Prokhor Thor}
\affiliation{\heriotwatt}

\author{Hyeonjun Baek}
\affiliation{Department of Physics, Sogang University, Seoul 04107, South Korea}

\author{Takashi Taniguchi}
\affiliation{\TsukubaTakashi}

\author{Kenji Watanabe}
\affiliation{\TsukubaKenji}

\author{Alejandro Molina-Sánchez}
\affiliation{\icmuv}

\author{Brian D. Gerardot}
\affiliation{\heriotwatt}

\author{Mauro Brotons-Gisbert}
\affiliation{\heriotwatt}
\email{m.brotons_i_gisbert@hw.ac.uk}

\date{\today}

\begin{abstract}
Moiré superlattices in transition-metal dichalcogenide semiconductor heterobilayers enable the quantum confinement of interlayer excitons with large out-of-plane permanent electric dipoles and spin-valley control. Here, we report a novel phonon-assisted excitation mechanism of individual moiré-trapped interlayer excitons in 2$H$-stacked MoSe$_2$/WSe$_2$ heterobilayers via chiral $E^{\prime\prime}$ in-plane optical phonons at the $\Gamma$-point. This excitation pathway preserves valley-selective optical selection rules and enables deterministic generation of individual interlayer excitons with defined helicity, emitting within a spectrally narrow energy spread. Through photoluminescence excitation spectroscopy in both the ensemble and quantum emitter regimes, we identify a fixed phonon energy of $\sim$23 meV mediating the process. First-principles calculations corroborate the symmetry and energy of the relevant phonon mode and its coupling to interlayer excitons, providing microscopic support for the observed valley-selective phonon-assisted excitation mechanism. Our results highlight the utility of chiral phonons as a tool for controlled excitation of quantum emitters in TMD moiré systems, opening new opportunities for valleytronic and quantum photonic applications.
\end{abstract}

\maketitle


\noindent Moiré superlattices formed by vertically stacking monolayers of transition-metal dichalcogenide (TMD) semiconductors have revolutionized condensed matter physics in recent years. By quenching the kinetic energy of charge carriers and creating a periodic potential landscape for both electrons and holes \cite{wu2018hubbard}—with lattice periods below 10 nm—these systems enter a regime where strong Coulomb interactions dominate, enabling the observation of a rich variety of strongly correlated electronic states \cite{mak2022semiconductor}. 
Beyond the confinement of electrons and holes into periodic superlattices, moiré lattices in TMD heterobilayers with type-II band alignment have also facilitated the trapping of interlayer excitons (IXs) \cite{yu2017moire,seyler2019signatures,brotons2020spin,baek2020highly,liu2021signatures,wang2021moire,brotons2021moire,baek2021optical}, which constitute the system’s lowest-energy Coulomb-bound electron–hole quasiparticles. 

While interlayer excitons (IXs) have been observed in various TMD heterobilayers \cite{fang2014strong,rivera2015observation,seyler2019signatures,jin2019observation,tran2019evidence,alexeev2019resonantly,tang2021tuning,zhao2024hybrid} and homobilayers \cite{gerber2019interlayer,leisgang2020giant,shimazaki2020strongly,peimyoo2021electrical,sponfeldner2021capacitively,feng2024highly}, quantum-dot-like confinement—evidenced by discrete emission lines from individual IXs—has, to date, only been reported in MoSe$_2$/WSe$_2$ systems \cite{seyler2019signatures,brotons2020spin,baek2020highly,liu2021signatures,wang2021moire,brotons2021moire,baek2021optical}. At low temperatures and under weak optical excitation, confocal photoluminescence (PL) spectra reveal a small number of sharp emission lines (with linewidths on the order of $\sim$100 $\mu$eV) typically spread over a broad energy range of up to $\sim$40 meV \cite{seyler2019signatures,brotons2020spin,baek2020highly,liu2021signatures,wang2021moire,brotons2021moire,baek2021optical}. These trapped IXs display strong helical polarization that reflects the underlying atomic registry and the crystal’s 
$C_3$ symmetry \cite{yu2017moire,yu2018brightened,seyler2019signatures,brotons2021moire}, and also exhibit uniform Landé g-factors, determined by the spin and valley configuration set by the relative twist angle between layers \cite{seyler2019signatures,baek2020highly,brotons2021moire}. The emission from a single moiré-confined IX shows photon antibunching \cite{baek2020highly}—a hallmark of its quantum emitter character—and can be tuned over a wide spectral range using an external vertical electric field \cite{baek2020highly}, thanks to its sizable out-of-plane permanent dipole moment. At higher excitation powers, the IX density increases, leading to broader emission features (up to $\sim$5 meV) and the emergence of multiple peaks corresponding to different IX species, such as charged excitons (trions), which can be controlled via gate voltages in tunable device architectures \cite{liu2021signatures,wang2021moire,brotons2021moire}. Moreover, to the best of our knowledge, moiré-trapped interlayer excitons in MoSe$_2$/WSe$_2$ heterobilayers remain the only excitonic species in TMDs and their heterostructures to exhibit single-photon emission while retaining the spin-valley physics, characteristic of the intralayer excitons in the constituent monolayers \cite{baek2021optical}. This effective locking of the optical selection rules to the spin, valley, and atomic registry of the layers allows for deterministic control of the emitted photon's helicity through the handedness of a circularly polarized pump laser, when resonant with an intralayer exciton in either layer \cite{wang2021moire,baek2020highly,baek2021optical}—typically requiring excitation energies 200–300 meV higher than those of the trapped IXs. 

However, a photoexcitation mechanism that simultaneously preserves spin--valley control of individual trapped interlayer excitons and selectively addresses a narrower subset of IXs has yet to be identified. In principle, resonance fluorescence would offer the most direct route to this regime, but its experimental implementation for moir\'e-trapped interlayer excitons remains extremely demanding and has not yet been achieved. Here, we demonstrate a phonon-assisted excitation pathway that bypasses these constraints, enabling deterministic, valley-selective excitation of individual moir\'e-trapped interlayer excitons within a narrow and tunable energy window.

Although a previous work reported the photoexcitation of IXs via exciton-phonon interaction \cite{shinokita2021resonant}, the large twist angle of the sample (12$^{\circ}$)—and therefore its small moiré period (2 nm)—prevented the trapping of individual IXs by the moiré lattice, which resulted in IX linewidths at least one order of magnitude broader than those reported for the single-photon emitters. In this work, we investigate the various photoexcitation pathways enabling quasi-resonant excitation of moiré-trapped IXs in twisted 2$H$-WSe$_2$/MoSe$_2$, in both the ensemble and quantum emitter regimes. Our results demonstrate that IXs can be efficiently generated via a phonon-assisted process involving a phonon with an energy of approximately 23 meV. The phonon energy, along with the optical selection rules governing both excitation and emission under resonant phonon-mode excitation, indicate that the phonon involved is the $E^{\prime\prime}$ chiral in-plane optical phonon at the $\Gamma$-point. First-principles calculations support this assignment by confirming the symmetry and energy of the $E^{\prime\prime}$ phonon and its coupling to interlayer excitons in the 2$H$-stacked heterobilayer. These $\Gamma$-point chiral phonons carry a well-defined pseudo-angular momentum that can be transferred to the excitonic system, effectively inverting the optical selection rules and resulting in IX emission with opposite helicity compared to resonant excitation of intralayer excitons in the individual monolayers. 
Altogether, our findings reveal that the fixed excess energy provided by the phonon-assisted process, combined with the phonon's chiral nature, enables a novel mechanism for the selective photoexcitation of single moiré-trapped IXs with defined chirality, emitting within a narrower energy window that can be efficiently tuned via the excitation laser energy.

\begin{figure}
    \centering
    \includegraphics[width=8.7cm]{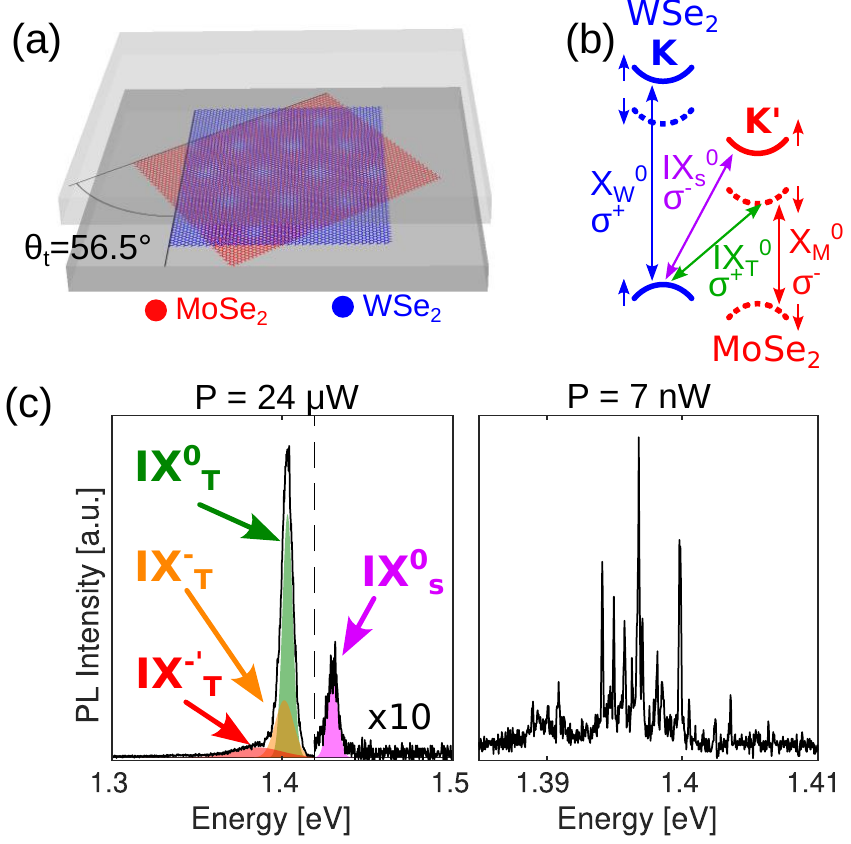}
    \caption{(a) Sketch of a twisted 2H-MoSe$_2$/WSe$_2$ heterostructure encapsulated by hBN. (b) Schematics of the type-II electronic band-edge alignment and spin-valley configuration in 2$H$-MoSe$_2$/WSe$_2$ at the corners of the hexagonal Brillouin zone corresponding to $K$ in WSe$_2$. The double-ended vertical blue and red arrows denote optical transitions corresponding to intralayer bright exciton states in WSe$_2$ and MoSe$_2$, respectively. The double-ended diagonal purple and green arrows represent optical transitions and optical selection rules arising from spin-singlet and spin-triplet neutral interlayer excitons pinned to a moiré site with an atomic registry $H^h_h$ \cite{yu2018brightened,seyler2019signatures,zhang2019highly,brotons2020spin,baek2020highly,brotons2021moire}. The helicity of the optical transitions is reversed for optical transitions at the opposite corner of the Brillouin zone. (c) PL spectra of interlayer excitons in the 2$H$-MoSe$_2$/WSe$_2$ sample under resonant excitation of the intralayer A exciton in monolayer MoSe$_2$ for excitation powers of  $P=24$ $\mu$W (left) and at $P=7$ nW (right). The different color-shaded Lorentzian peaks indicate ensemble PL emission from different interlayer exciton resonances.}
    \label{fig:Recombination_mechanism}
\end{figure}

Figure \ref{fig:Recombination_mechanism}(a) shows a sketch of the stacked 2$H$-type MoSe$_2$/WSe$_2$ heterobilayer employed in our study. The  MoSe$_2$/WSe$_2$ heterobilayer is encapsulated by hexagonal boron nitride (h-BN). Details of sample fabrication can be found in Ref. \cite{brotons2021moire}. A relative twist angle $\theta_t=56.5\pm 0.8^\circ$ was determined by polarization-resolved second harmonic generation (see Supplementary Information). The measured twist angle is above the theoretically proposed critical angle for lattice reconstruction \cite{rosenberger2020twist,weston2020atomic,andersen2021excitons}, ensuring minimal domain formation \cite{zhao2023excitons}. Figure \ref{fig:Recombination_mechanism}(b) shows a schematics of the type-II electronic band-edge alignment and spin-valley configuration in 2$H$-MoSe$_2$/WSe$_2$, which results in a minimum displacement in momentum space between the carriers at the $K$ and $K'$ valleys in WSe$_2$ and MoSe$_2$. The double-ended vertical blue and red arrows denote optical transitions corresponding to intralayer bright exciton states in WSe$_2$ (X$_W^0$) and MoSe$_2$ (X$_M^0$), respectively. The double-ended diagonal purple and green arrows represent optical transitions arising from spin-conserving (i.e., spin-singlet IX$_S^0$) and spin-flipping (i.e., spin-triplet IX$_T^0$) neutral interlayer excitons in 2$H$-MoSe$_2$/WSe$_2$ \cite{yu2018brightened,seyler2019signatures,zhang2019highly,brotons2020spin,baek2020highly,brotons2021moire}. The selection rules corresponding to each optical transition are also indicated in the schematics. Note that the selection rules of the interlayer optical transitions correspond to those of spin-singlet and spin-triplet interlayer excitons spatially pinned by moiré trapping sites with an atomic interlayer registry $H_h^h$, in agreement with the theoretical predictions and previous experimental results in this system \cite{yu2018brightened,seyler2019signatures,zhang2019highly,brotons2020spin,baek2020highly,brotons2021moire}. Here, $H_h^h$ denotes the $H$-type stacking in which the hexagon centres ($h$) of the two layers are vertically aligned. Low-temperature (T$\sim$ 4 K) confocal photoluminescence (PL) measurements reveal the presence of moiré-trapped interlayer exciton emission, in agreement with previous studies in this sample \cite{baek2020highly,brotons2021moire,baek2021optical} and similar 2$H$-stacked MoSe$_2$/WSe$_2$ heterobilayers \cite{seyler2019signatures,liu2021signatures,wang2021moire}. 

Figure \ref{fig:Recombination_mechanism}(c) shows IX PL spectra measured at a representative spot of our heterobilayer under two different excitation power regimes. A continuous-wave (CW) excitation laser was used to resonantly excite the 1$s$ state of the intralayer A exciton in monolayer MoSe$_2$ ($\lambda\sim759$ nm). At low excitation powers ($P\approx7$ nW) the PL spectrum shows several discrete narrow emission lines with energies in the range 1.39–1.41 eV (right panel of Fig. \ref{fig:Recombination_mechanism}(c)), a signature of IX$_T^0$ trapped in the moiré potential landscape \cite{seyler2019signatures,baek2020highly,liu2021signatures,wang2021moire,brotons2021moire,baek2021optical}. The IX$_T^0$ nature of the trapped IXs was confirmed by their Landé $g$-factor (see Supplementary Information), a clear indicator of both the relative valley alignment between the layers and the spin configuration of the carriers. We note that a recent experimental study on a twisted 2$H$-MoSe$_2$/WSe$_2$ heterobilayer has reported the observation of uniformly spaced emission lines, which the authors attributed to phonon replicas of the IX, specifically involving a phonon energy of 800 $\mu$eV, rather than to moiré-induced confinement \cite{soubelet2024polarons}. In contrast, our spectrally narrower emission lines do not exhibit any clear uniform energy spacing and frequently show energy separations below 200 $\mu$eV. At higher excitation powers of $P\approx24$ $\mu$W (left panel of Fig. \ref{fig:Recombination_mechanism}(c)), the spectrum shows a broader IX ensemble PL peak that preserves the magneto-optical properties of the narrow IX$_T^0$ emission lines \cite{brotons2021moire}.  The low-energy tail of the ensemble PL peak deviates from a Lorentzian lineshape due to the presence of three different excitonic resonances, as indicated by our multi-Lorentzian fit (see Supplementary Information). The brightest excitonic PL resonance (green-shaded Lorentzian) corresponds to the ensemble IX$_T^0$ peak. A resonance corresponding to the intervalley IX trion ensemble peak with spin-triplet optical configuration (IX$^{-}_{T}$, orange-shaded peak) can be observed approximately 7 meV below the IX$_T^0$ ensemble PL peak (see Supplementary Material). A third excitonic resonance labeled IX$^{-'}_{T}$ (red-shaded Lorentzian), whose origin is not yet fully understood \cite{brotons2021moire}, can also be observed at lower energy than IX$^{-}_{T}$. The presence and binding energy of IX$^{-}_{T}$ are consistent with previously published results that demonstrated that the band alignment in 2$H$-MoSe$_2$/WSe$_2$ heterobilayers allows the formation of IX$^{-}_{T}$ by optical pumping even in the neutral doping regime \cite{brotons2021moire}. Finally, the spectrum further shows the spin-singlet IX$_S^0$ ensemble PL peak (pink-shaded Lorentzian) at energies approximately 25 meV above the ground spin-triplet IX$_T^0$, in agreement with previous observations \cite{jauregui2019electrical,joe2021electrically,brotons2021moire}.


\begin{figure*}
    \centering
    \includegraphics[width=18cm]{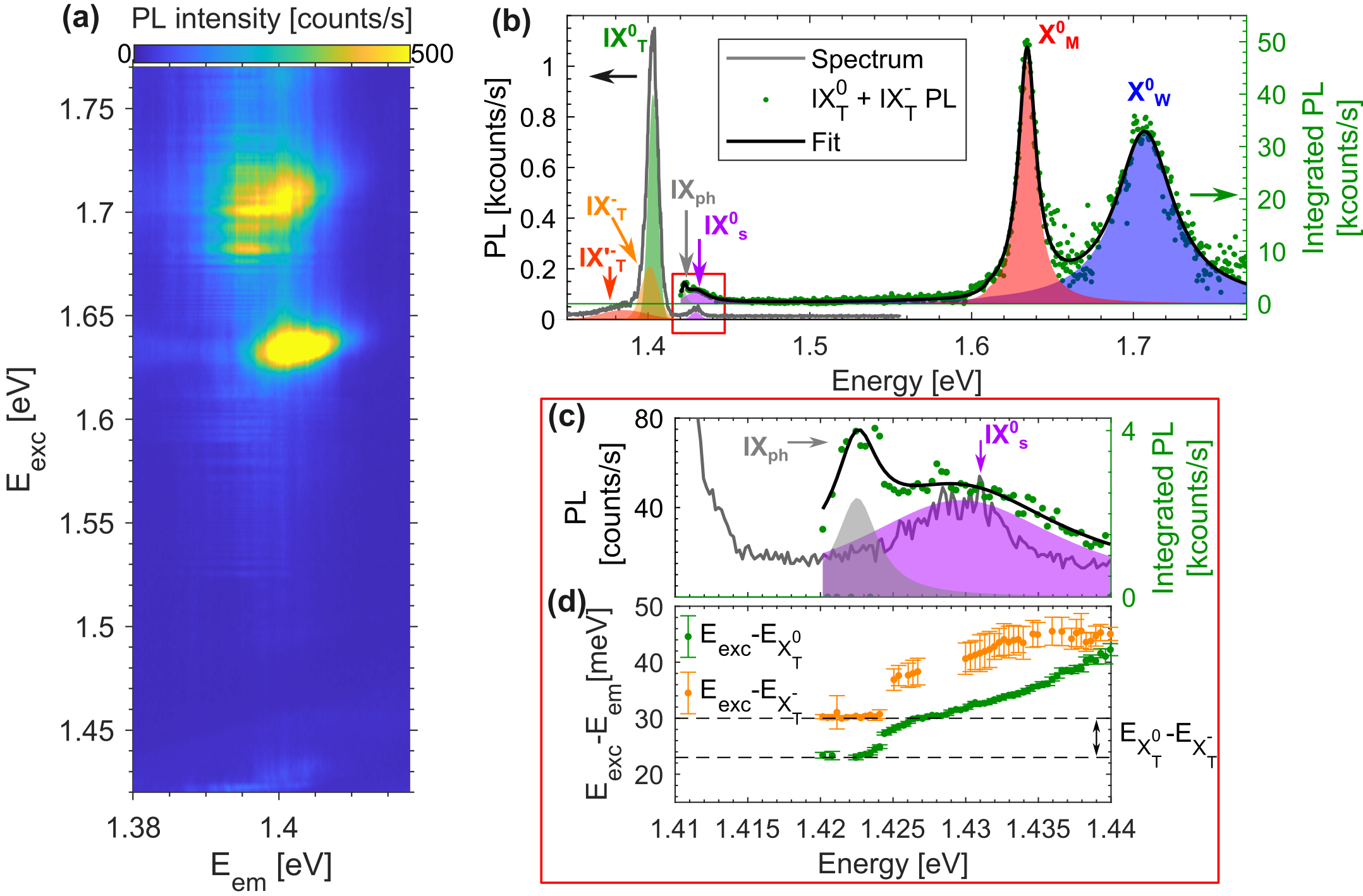}
    \caption{(a) False colour-plot of the IX PL spectra recorded with respect to the CW-laser excitation energy ($E_{exc}$ under a constant power of 25 $\pm$ 3 $\mu$W on the sample. (b) Representative IX PL spectrum (gray line, left axis) superposed with the combined integrated PL intensity of the IX$^0_T$ and IX$^-_T$ ensemble peaks as a function of excitation energy (green dots, right axis). The black solid line represents a four-Lorentzian-peak fit of the integrated PL intensity.
    (c) Zoom in of panel (b) in the energy range 1.41 - 1.44 eV, as indicated by the red square in (b). (d) Excess energies measured for IX$^0_T$ and IX$^-_T$ as a function of E$_{exc}$.}
    \label{fig:PLE}
\end{figure*}

To investigate the different photoexcitation paths of moir{\'e}-trapped IXs, we carried out photoluminescence excitation resolved (PLE) measurements. Figure \ref{fig:PLE}(a) shows a false colour-plot of the IX PL spectra recorded with respect to the CW laser excitation energy under a constant power of 25 $\pm$ 3 $\mu$W on the sample. Figure \ref{fig:PLE}(b) shows a representative IX PL spectrum (gray line). As discussed in Fig. \ref{fig:Recombination_mechanism}(c), we observe the presence of different excitonic resonances corresponding to IX$^0_T$, IX$^0_S$, IX$^-_T$, and IX$^{-'}_T$. Figure \ref{fig:PLE}(a) further shows that the photoexcitation of IXs is enhanced for excitation energies of $\sim1.64$ eV and $\sim1.71$ eV, which can be attributed to the efficient photon absorption at energies resonant with the ground (1$s$) state intralayer A excitons in MoSe$_2$ (X$^0_M$) and WSe$_2$ (X$^0_W$), respectively \cite{seyler2019signatures,baek2020highly}. Two dimmer resonances can also be seen for excitation energies $\sim1.42-1.44$ eV. Interestingly, we note that the IX$^0_T$ peak is predominant for excitation energies above 1.424 eV, while the IX$^-_T$ peak becomes more intense for laser energies below 1.424 eV (see Supplementary Information).

To explore these low-energy excitation resonances in more detail, we plot the combined integrated PL intensity of the IX$^0_T$ and IX$^-_T$ ensemble peaks as a function of the excitation energy (green dots in Fig. \ref{fig:PLE}(b)). As observed in Fig. \ref{fig:PLE}(a), the integrated PL intensity displays four Lorentzian-shaped absorption resonances: two resonances at high energy corresponding to X$^0_M$ and X$^0_W$, and two weaker resonances below 1.45 eV. Figure \ref{fig:PLE}(c) shows a zoom-in of Fig. \ref{fig:PLE}(b) in the energy range $1.41-1.44$ (as indicated by the red box in Fig. \ref{fig:PLE}(b)). We observe that the resonance at $\sim$1.43 eV overlaps in energy with the IX$^0_S$ PL emission peak, indicating an enhanced photoexcitation of IX$^0_T$ via resonant excitation of the higher energy IX$^0_S$ state (see band diagram in Fig. \ref{fig:Recombination_mechanism}(b)). We therefore label this PLE peak as IX$^0_S$. Note that photoexcitation of IX$^0_T$ via absorption by the IX$^0_S$ state is less efficient at generating IX$^0_T$ than direct resonant excitation of the intralayer excitons. This is due to the reduced oscillator strength of the interlayer IX$^0_S$ exciton. The PLE spectrum also reveals a narrower absorption resonance, with a full width at half maximum (FWHM) of approximately 4 meV, centered at an energy $\sim$7 meV below the IX$_S^0$ resonance. A similar PLE resonance, labeled IX$_{ph}$, has previously been observed and attributed to a phonon-assisted emission process \cite{shinokita2021resonant}.

A constant excess energy between excitation and emission is a hallmark of single-phonon–assisted absorption. To corroborate the phonon-assisted nature of the IX$_{ph}$ resonance, we extract the excess energy $E_{exc}-E_{em}$, defined as the difference between the excitation ($E_{exc}$) and emission ($E_{em}$) energies. In a single phonon-assisted process, energy conservation dictates $E_{exc}-E_{em}=E_{ph}$, where $E_{ph}$ represents the phonon energy. Therefore, a key signature of phonon-assisted processes is a constant excess energy for any $E_{exc}$ within the excitonic resonance. Figure \ref{fig:PLE}(d) presents the excess energies measured for IX$^0_T$ (E$_{exc}-E_{IX_T^0}$) and IX$^-_T$ (E$_{exc}-E_{IX^-_T}$) as a function of E$_{exc}$, with $E_{IX_T^0}$ ($E_{IX^-_T}$) the energy of the IX$^0_T$ (IX$^-_T$) peak. The evolution of excess energy for IX$^0_T$ and IX$^-_T$ exhibits a linear dependence on $E_{exc}$ for values outside the IX$_{ph}$ resonance, which confirms that the IX$_S^0$ peak in the PLE spectrum originates from resonant excitation of the spin-singlet interlayer exciton transition. Interestingly, for $E_{exc}$ within the IX$_{ph}$ peak, the excess energy of IX$^0_T$ and IX$^-_T$ plateaus at constant values of approximately 23 meV and 30 meV, respectively. This observation corroborates the phonon-assisted nature of the IX$_{ph}$ resonance, and suggests the involvement of a phonon process with an energy of $\sim$23 meV. We note that the difference in the excess energy at the plateaus between the IX$^0_T$ and IX$^-_T$ arises from the binding energy of the IX$^-_T$ state (i.e., $E_{X_T^0}-E_{X_T^-}\approx7$ meV \cite{brotons2021moire,baek2021optical}).

The extracted phonon energy of $\sim$23 meV (185.5 cm$^{-1}$) is in good agreement with the phonon excess energy of 24 meV observed in the PLE spectrum of a MoSe$_2$/WSe$_2$ heterobilayer reported in Ref. \cite{shinokita2021resonant}, which was attributed to an $A_1^\prime$ phonon mode in either MoSe$_2$ or WSe$_2$ (typically around 240 cm$^{-1}$). A previous work on MoSe$_2$/WSe$_2$ heterobilayers has also reported the observation of two phonon resonances at 22.3 meV (179.9 cm$^{-1}$) and 22.95 meV (185.1 cm$^{-1}$) using the magneto-phonon resonance effect \cite{delhomme2020flipping}. In this case, the authors attributed the origin of the observed phononic resonances to chiral in-plane $E^{\prime\prime}$ optical phonons at the zone centre ($\Gamma$) of the Brillouin zone in MoSe$_2$ and WSe$_2$, with energies $\sim$21-22 meV \cite{soubelet2016resonance,luo2013effects,chen2019entanglement}. The $E^{\prime\prime}$ optical phonon modes in MoSe$_2$ and WSe$_2$ are non-Raman-active \cite{delhomme2020flipping}, which agrees with our room-temperature unpolarized Raman spectroscopy results (see Supplementary Material). The Raman spectrum of the heterobilayer exhibits clear peaks at 243.8 cm$^{-1}$ ($\sim$30 meV) and 252 cm$^{-1}$ ($\sim$31 meV) corresponding to $A_{1g}$ phonon modes in MoSe$_2$ and WSe$_2$, respectively. In contrast, no Raman signature is observed at $\sim$185 cm$^{-1}$. The large difference in the excess energy of IX$_{ph}$ (23 meV) and the $A_{1g}$ phonon modes ($\sim$30 meV), together with the Raman inactivity of the phonon mode responsible for IX$_{ph}$ at $\sim$185 cm$^{-1}$, allows us to exclude the $A_{1g}$ phonon modes as the origin of IX$_{ph}$.

The measured phonon energy of approximately 23 meV in our work, as well as those reported in previous studies (22-24 meV) \cite{delhomme2020flipping,shinokita2021resonant}, can therefore be attributed to in-plane chiral $E^{\prime\prime}$ optical phonons \cite{delhomme2020flipping}. The extracted phonon energies align well with first-principles calculations for these in-plane $E^{\prime\prime}$ modes, whether at the zone center ($\Gamma$) or at the corners ($K$ and $K$') of the hexagonal Brillouin zone in individual monolayers \cite{chen2019entanglement,mahrouche2022phonons}. Notably, a similar agreement is observed in first-principles calculations results for MoSe$_2$/WSe$_2$ heterobilayers, which have been shown to preserve the intrinsic phonon characteristics of their free-standing monolayer constituents \cite{mahrouche2022phonons}. The chirality of these phonon modes at the high-symmetry points in the center and corners of the Brillouin zone has a different origin. At $\Gamma$, the superposition of the energy-degenerate linearly polarized longitudinal optical (LO) and transverse optical (TO) $E^{\prime\prime}$ modes results in a circular motion of the atoms \cite{zhang2015chiral,chen2019entanglement}. At the $K$ and $K^\prime$ zone corners, the combination of broken inversion symmetry of the lattice and time-reversal symmetry lifts the degeneracy of the clockwise and counterclockwise $E^{\prime\prime}$ phonon modes, resulting in an intrinsic phonon chirality that is opposite between the two valleys \cite{zhang2015chiral,zhu2018observation}.

\begin{figure}
    \centering
    \includegraphics[width=8cm]{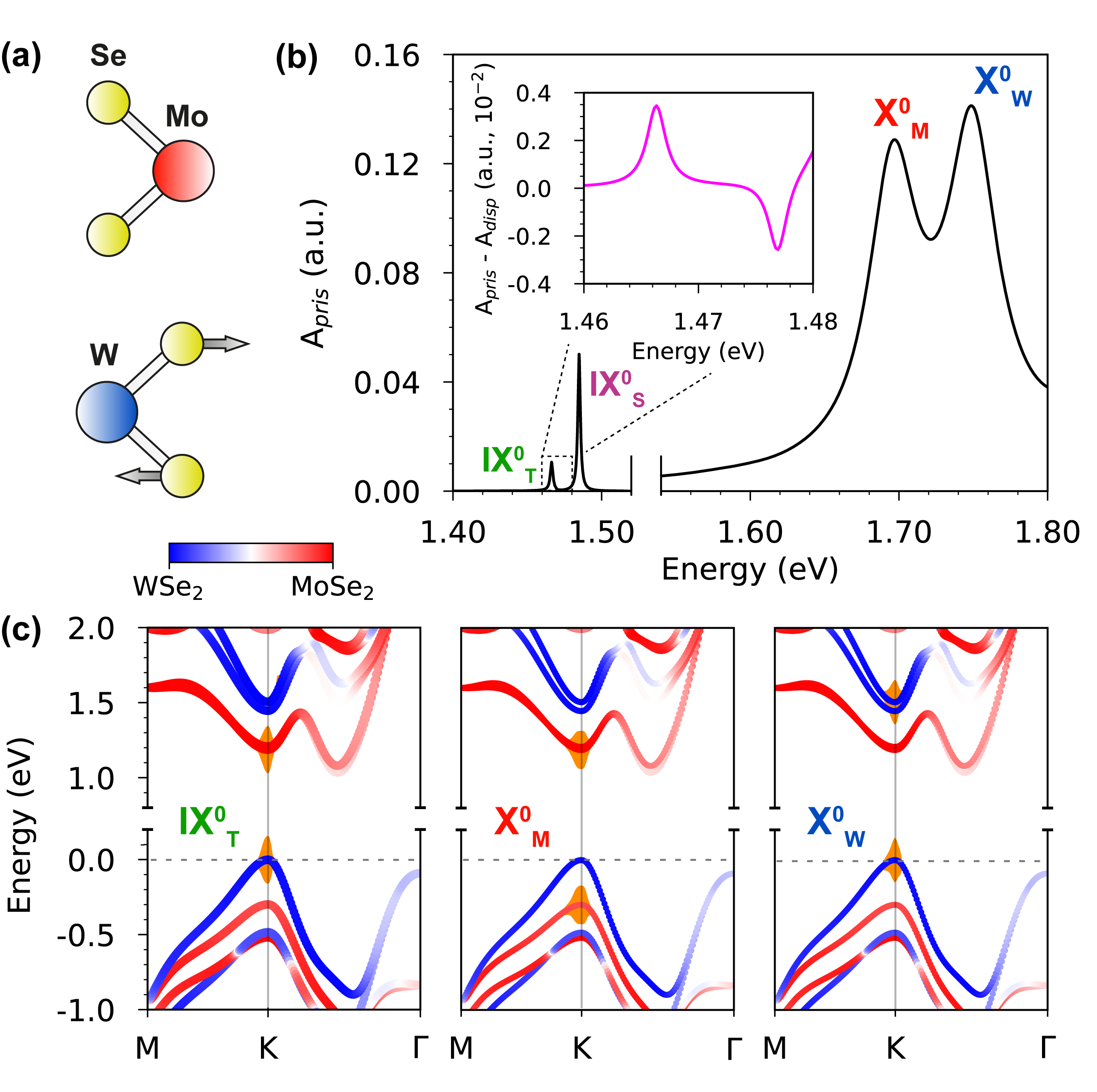}
    \caption{(a) Schematic representation of the $E^{\prime\prime}$ phonon mode in the MoSe$_{2}$/WSe$_{2}$ heterobilayer. The arrows indicate the displacement pattern of the Se atoms associated with this vibrational mode. (b) Absorbance spectrum of the pristine MoSe$_{2}$/WSe$_{2}$ heterobilayer, with the main excitonic resonances labeled. A broadening of 0.03 eV was employed for the intralayer excitons, while a smaller broadening of 0.001 eV was used for the interlayer excitons in order to properly resolve their weak absorption features. The inset displays the differential absorbance between the pristine structure and the configuration distorted according to the $E^{\prime\prime}$ phonon eigenvector. A pronounced variation at the IX$^{0}_{T}$ peak reveals strong exciton--phonon coupling between the interlayer exciton and this phonon mode. (c) Illustration of the different excitonic states (orange area) mapped onto the layer-resolved band structure, confirming the origin and composition of each labeled exciton.}
    \label{fig:theory}
\end{figure}

To support the interpretation of the IX$_{ph}$ resonance as originating from a phonon-assisted process involving the spin-triplet IX$^{0}_{T}$ and the $E^{\prime\prime}$ phonon mode, we performed state-of-the-art first-principles calculations (see Supplementary Material for computational details). Phonon energies were computed at the $\Gamma$ point of the MoSe$_2$/WSe$_2$ heterobilayer in the AA$^{\prime}$ stacking configuration, yielding an energy of 21.14~meV for the $E^{\prime\prime}$ mode. This value is in good agreement with both our experimental observations and previous theoretical reports \cite{mahrouche2022phonons}. A schematic representation of the atomic displacement pattern associated with the $E^{\prime\prime}$ phonon is shown in Fig.~\ref{fig:theory}(a).

The calculated absorbance spectrum of the pristine heterobilayer is presented in Fig.~\ref{fig:theory}(b), where the dominant excitonic resonances are labeled. According to spin-conserving optical selection rules, the spin-triplet IX$^{0}_{T}$ is expected to be optically dark, as it originates from a spin-forbidden transition and therefore should not contribute to the linear absorption. However, in the MoSe$_2$/WSe$_2$ heterobilayer the absence of out-of-plane mirror symmetry enables a mixing of spin-allowed and spin-forbidden excitonic states mediated by spin--orbit coupling, partially relaxing the optical selection rules of the monolayer\cite{yu2018brightened}. This symmetry breaking renders the IX$^{0}_{T}$ weakly optically active, allowing it to appear as a finite feature in the absorption spectrum.

To accurately describe both intralayer and interlayer excitons, distinct broadening parameters were employed in the calculation of the absorbance. The interlayer or intralayer character of the relevant excitonic resonances was further confirmed by projecting their contributions (orange regions) onto the layer-resolved electronic band structure shown in Fig.~\ref{fig:theory}(c), yielding a classification consistent with the experimental assignment.

To assess the role of the $E^{\prime\prime}$ phonon mode in the emergence of the IX$_{ph}$ resonance, we computed the absorbance spectrum of a structure statically distorted along the $E^{\prime\prime}$ phonon eigenvector shown in Fig.~\ref{fig:theory}(a). Atomic displacements corresponding to 1\% of the W--Se bond length were applied in order to remain within the linear regime. The difference between the absorbance spectra of the pristine and distorted structures, shown in the inset of Fig.~\ref{fig:theory}(b), exhibits a pronounced modification at the energy of the IX$^{0}_{T}$ exciton. This finding provides clear evidence of a non-negligible exciton--phonon coupling between the $E^{\prime\prime}$ phonon mode and the IX$^{0}_{T}$ exciton, supporting the scenario in which the $E^{\prime\prime}$ phonon actively participates in the formation of the IX$_{ph}$ resonance.

\begin{figure}
    \centering
    \includegraphics[width=8.7cm]{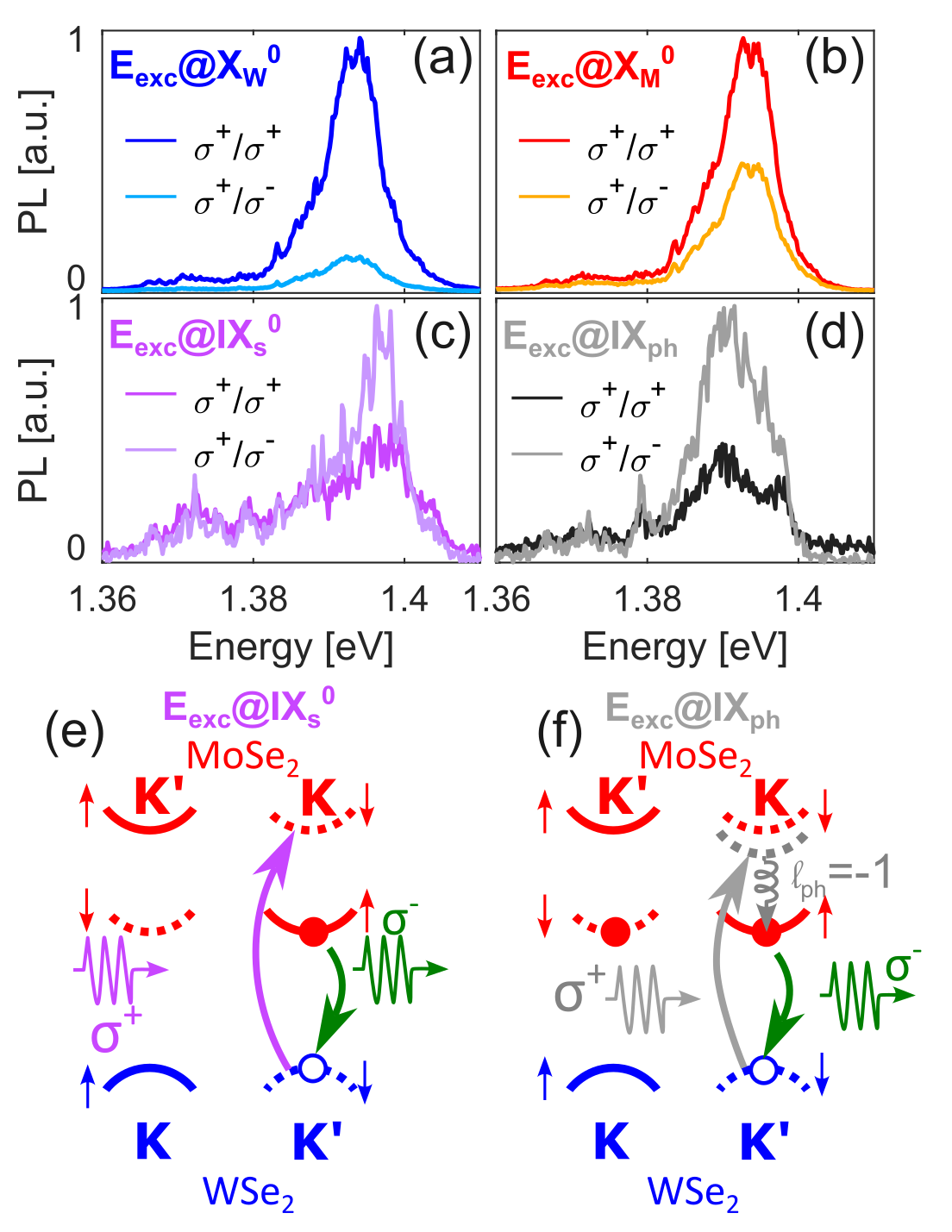}
    \caption{Helicity-resolved photoluminescence spectra of interlayer excitons in the ensemble regime under different resonant excitation conditions. (a)–(c) Circularly polarized PL spectra recorded under $\sigma^{+}$ excitation resonant with the WSe$_2$ intralayer exciton X$_W^0$ (a), the MoSe$_2$ intralayer exciton X$_M^0$ (b), and the interlayer singlet exciton IX$_S^0$ (c), respectively. Panels (a)–(c) illustrate excitation pathways that do not involve phonon angular momentum transfer and therefore follow the optical selection rules expected for direct electronic transitions in 2H-MoSe$_2$/WSe$_2$. (d) Helicity-resolved PL spectrum under excitation resonant with IX$_{ph}$, corresponding to chiral-phonon–assisted excitation. The resulting cross-circularly polarized emission reflects the transfer of pseudo-angular momentum from the $\Gamma$-point chiral phonon to the excitonic system. (e),(f) Schematic illustrations of the excitation–recombination pathways for resonant excitation of IX$_S^0$ (e) and phonon-assisted excitation of IX$_{ph}$ leading to IX$_T^{-}$ formation (f).} 

    \label{fig:SelectionRules}
\end{figure}

Interestingly, the preservation of rotational $C_3$ symmetry at the $\Gamma$ high-symmetry point endows the in-plane chiral $E^{\prime\prime}$ optical phonons with a quantized pseudo-angular momentum (PAM) along the out-of-plane direction ($l_{ph}$) \cite{zhang2015chiral,zhu2018observation,chen2019entanglement,delhomme2020flipping}. Specifically, the degenerate TO+LO mode at $\Gamma$ carries a PAM $l_{ph}=\pm1$ \cite{zhang2015chiral}. Furthermore, the conservation of PAM directly influences the selection rules governing optical transitions involving chiral phonons \cite{zhu2018observation,delhomme2020flipping}. The fundamental conservation laws of energy ($E$), momentum ($k$), and angular momentum ($l$) impose the following selection rules on these transitions:  
\begin{equation} 
    \begin{split}
     E_c-E_v & =  E_{\gamma} + E_{ph}  \\
     k_c-k_v & = k_{\gamma} + k_{ph} \\
     l_c-l_v & = l_{\gamma} + l_{ph}, \\
    \end{split}
    \label{eq1:SelectionRules}
\end{equation}
where the subscripts $c$ and $v$ denote the conduction and valence bands, respectively, while $\gamma$ and $ph$ refer to the photon and phonon transitions. Therefore, optical selection rules can provide additional information about the nature of the phonon involved. We begin by exploring the optical selection rules of IX$^0_T$ and IX$^-_T$ in the ensemble regime under different photoexcitation mechanisms. Figures \ref{fig:SelectionRules}(a)-(d) show helicity-resolved IX PL spectra under resonant excitation of the four resonances identified in the PLE spectrum of Fig. \ref{fig:PLE}(b): X$_W^0$ (a), X$_M^0$ (b), IX$_S^0$ (c), and IX$_{ph}$ (d). We set the excitation polarization to be circularly polarized with helicity $\sigma^+$ and detected either the $\sigma^+$ (co-polarized) or $\sigma^-$ (cross-polarized) photon emission. The photon emission by IXs under resonant excitation of X$_W^0$, X$_M^0$, and IX$_S^0$ does not involve a phonon process (i.e., $E_{ph}=k_{ph}=l_{ph}=0$). For these photoexcitation paths, we therefore observe the optical selection rules predicted theoretically for 2$H$-MoSe$_2$/WSe$_2$ from the symmetry analysis of the conduction and valence band states involved in the absorption and emission processes \cite{yu2018brightened,gilardoni2021symmetry}, which agree with previous experimental observations \cite{seyler2019signatures,baek2020highly,brotons2021moire}. Resonant excitation of X$_W^0$ and X$_M^0$ leads to co-circularly polarized IX$^0_T$ emission, whereas resonant excitation of IX$^0_S$ results in cross-circularly polarized emission. Note that some depolarization occurs due to long-range electron-hole exchange within the individual monolayers \cite{yu2014valley,glazov2014exciton,yang2020exciton}, with spin depolarization being faster in MoSe$_2$ \cite{yang2020exciton}.  Figure \ref{fig:SelectionRules}(e) shows a schematic of the photoexcitation and recombination process of IX$^0_T$ via resonant excitation of IX$^0_S$. The cross-circularly polarized optical selection rules of IX$^0_S$ and IX$^0_T$ lead to opposite circular polarization for the photon absorption and emission processes \cite{yu2018brightened}. 

\begin{figure*}
    \centering
    \includegraphics[width=15cm]{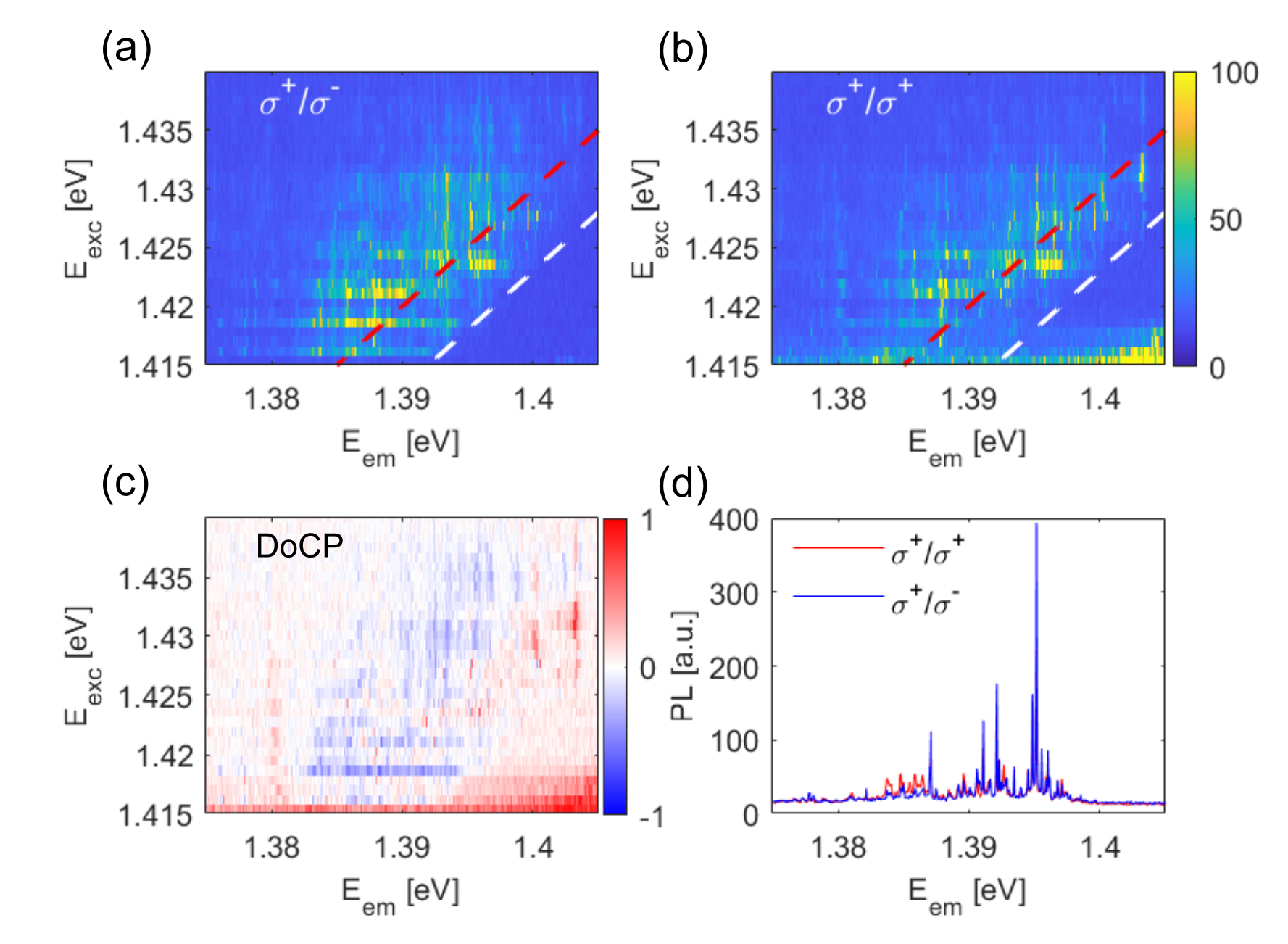}
    \caption{Excitation resolved photoluminescence resolved in polarization. Excitation is $\sigma^+$ circularly polarized and PL is detected in either $\sigma^-$ (a) or $\sigma^+$ (b) polarization. Red dotted line corresponds to $E_{exc}-$30 meV and the white dotted line corresponds to $E_{exc}-$23 meV. Single emitter energies follow better the red dotted line indicating the formation of trions. (c) Valley polarization as a function of the excitation energy. (d) Example of single emitter spectrum for $E_{exc}=1.4226$ eV.}
    \label{fig:DoCP_vs_power}
\end{figure*}

Notably, resonant excitation of IX$_{ph}$ results in cross-circularly polarized photon emission from IX$_T^{-}$, which is the predominant IX species under excitation energies resonant with IX$_{ph}$ (see previous discussion of the PLE results in Fig. \ref{fig:PLE}). Within a single-particle picture, the charge configuration of the intervalley IX$_T^{-}$ involves one electron occupying the lowest conduction band in both the $K$ and $K^{\prime}$ valleys (see schematics in Fig. \ref{fig:SelectionRules}(f)). The observed $\sigma^{-}$ photon emission polarization from IX$_T^{-}$ indicates radiative recombination between an electron at $K$ in MoSe$_2$ and a hole at $K^{\prime}$ in WSe$_2$. However, it should be noted that, in general, $\sigma^+$-excitation leads to the creation of electron-hole pairs in the $K$ valley of WSe$_2$ \cite{xiao2012coupled}. Therefore, phonon-assisted photoexcitation of IX$_T^{-}$ via resonant excitation of IX$_{ph}$ necessarily involves a hole state at the $K^{\prime}$ valley in the top valence band of WSe$_2$. This behavior can be explained by the emission of a phonon with momentum $k_{ph} = 0$, angular momentum $l_{ph} = -1$, and energy $E_{ph} \approx 23$ meV—characteristics consistent with the $E^{\prime\prime}$ chiral in-plane optical phonons at the $\Gamma$-point, in agreement with the findings of Ref. \cite{delhomme2020flipping}. We note that $l_{ph} = -1$ is equivalent to $l_{ph} = +2$ due to the $C_3$ symmetry of the lattice, where angular momentum is defined modulo 3 \cite{delhomme2020flipping}.


Next, we explore the phonon-assisted excitation of moiré-trapped IXs via chiral phonons in the quantum emitter regime. Figures \ref{fig:DoCP_vs_power}(a) and \ref{fig:DoCP_vs_power}(b) show PLE measurements of individual moiré-trapped IXs performed at an excitation power of $P_{\text{exc}} \sim$ 200 nW, with excitation energies scanned across the IX$_{ph}$ and IX$_S^{0}$ resonances, under cross- (a) and co-polarized (b) circular excitation/detection configurations. We observe a linear correlation between the excitation energy ($E_{\text{exc}}$) and the emission energy ($E_{\text{em}}$) of the individual emitters, confirming the phonon-assisted creation of moiré-trapped IXs. The diagonal red and white dashed lines in Figs. \ref{fig:DoCP_vs_power}(a) and \ref{fig:DoCP_vs_power}(b) indicate constant excess energies of 30 meV and 23 meV, respectively. As shown, the trapped excitons exhibit enhanced PL intensity at an excess energy of approximately 30 meV. This observation is consistent with the extracted phonon energy of $\sim$23 meV and the additional trion binding energy of roughly 7 meV \cite{wang2021moire,liu2021signatures,baek2021optical}, and provides a mechanism to selectively photoexcite single
moiré-trapped emitting within
a given energy window. Further, we carry a power-dependent study of the moiré-trapped IXs under resonant excitation of IX$_{ph}$, which corroborates their trion nature (see Supplementary Material). Finally, similar to the results obtained in the IX ensemble regime, resonant excitation of IX$_{ph}$ leads to cross-circularly polarized photon emission from IX$_T^{-}$, as shown in Figs. \ref{fig:DoCP_vs_power}(c) and \ref{fig:DoCP_vs_power}(d). Taken together, these results support the interpretation that single moiré-trapped IXs can be photoexcited via the absorption of chiral in-plane optical phonons at the $\Gamma$-point.


In summary, we investigated the various photoexcitation pathways enabling quasi-resonant excitation of moiré-trapped IXs in twisted 2$H$-WSe$_2$/MoSe$_2$, in both the ensemble and quantum-emitter regimes. Our results demonstrate that, in addition to the direct photoexcitation of IXs via resonant excitation of the higher-energy spin-singlet IX state, IXs can also be generated through a phonon-assisted absorption process involving a phonon with an energy of approximately 23 meV. The phonon energy, together with the optical selection rules governing both excitation and emission under resonant phonon-mode excitation, identifies the phonon involved as the chiral $E^{\prime\prime}$ in-plane optical mode at the $\Gamma$ point. First-principles calculations provide microscopic support for this interpretation by confirming the symmetry and energy of the $E^{\prime\prime}$ phonon and its coupling to interlayer excitons in the 2$H$-stacked heterobilayer. These $\Gamma$-point chiral phonons can transfer their angular momentum, resulting in IX emission with optical selection rules opposite to those enabled by resonant excitation of intralayer excitons in the individual monolayers. Importantly, the phonon-assisted pathway demonstrated here provides many of the advantages of resonance fluorescence—selectivity, helicity control, and spectral narrowing—without the stringent experimental requirements associated with strict resonant excitation.
Altogether, our findings reveal that the fixed excess energy provided by the phonon-assisted process, combined with the phonon’s chiral nature, enables a novel mechanism for the selective photoexcitation of single moiré-trapped IXs with defined chirality, emitting within an energy window that can be efficiently tuned via the excitation laser energy.\\

\section*{ACKNOWLEDGEMENTS}
This work was supported by the EPSRC (grant nos. EP/P029892/1 and EP/Y026284/1). M.B.-G. is supported by a Royal Society University Research Fellowship. B.D.G. is supported by a Chair in Emerging Technology from the Royal Academy of Engineering. K.W. and T.T. acknowledge the support from the JSPS KAKENHI (Grant Numbers 21H05233 and 23H02052) , the CREST (JPMJCR24A5), JST and World Premier International Research Center Initiative (WPI), MEXT, Japan. This work is supported by the Horizon Europe research and innovation program of the European Union under the Marie Sklodowska-Curie grant agreement 101118915 (TIMES). This work is part of the project I+D+i PID2023-146181OB-I00 UTOPIA, funded by MCIN/AEI/10.13039/501100011033, the project PROMETEO/2024/4 (EXODOS) and SEJIGENT/2021/034 (2D-MAGNONICS) funded by the Generalitat Valenciana. This study is also part of the Advanced Materials program (project SPINO2D), supported by MCIN with funding from the European Union NextGenerationEU (PRTR-C17.I1) and Generalitat Valenciana. J.C.V and A.M.S. thankfully acknowledge the computer resources at Agustina and technical support provided by BIFI and Barcelona Supercomputing Center (FI-2025-2-0001), and the computer resources at Tirant-UV (project lv48 - FI-2025-2-0001). H. B. acknowledges the support from the POSCO Science Fellowship of POSCO TJ Park Foundation and the National Research Foundation (NRF) of Korea grants funded by the Korean Ministry of Science and ICT (Grant Nos. RS-2023-00220471, RS-2025-24683194, RS-2025-14532973, and RS-2025-25463680)

\bibliography{Bibliography}

\end{document}